\documentclass{emulateapj}
\usepackage{amsmath}
\usepackage{amssymb}
\usepackage{epsf}

\shorttitle{The Structure of the Homunculus}
\shortauthors{Smith \& Townsend}
\begin{document}

\title{The Structure of the Homunculus. III. Forming a Disk and
  Bipolar Lobes in a Rotating Surface Explosion\altaffilmark{1}}

\author{Nathan Smith} \affil{Astronomy Department, 601 Campbell Hall,
University of California, Berkeley CA 94720; nathans@astro.berkeley.edu}

\and 

\author{Richard H.D.\ Townsend} \affil{Bartol Research Institute,
University of Delaware, Newark, DE 19716; rhdt@batol.udel.edu}

\altaffiltext{1}{Based in part on observations obtained at the Gemini
Observatory, which is operated by AURA, under a cooperative agreement
with the NSF on behalf of the Gemini partnership: the National Science
Foundation (US), the Particle Physics and Astronomy Research Council
(UK), the National Research Council (Canada), CONICYT (Chile), the
Australian Research Council (Australia), CNPq (Brazil), and CONICET
(Argentina).}

\begin{abstract}

We present a semi-analytic model for the shaping of the Homunculus
Nebula around $\eta$~Carinae that accounts for the simultaneous
production of bipolar lobes and an equatorial disk through a rotating
surface explosion.  Material is launched normal to the surface of an
oblate rotating star with an initial kick velocity that scales
approximately with the local escape speed.  Thereafter, ejecta follow
ballistic orbital trajectories, feeling only a central force
corresponding to a radiatively reduced gravity. Our model is
conceptually similar to the wind-compressed disk model of Bjorkman \&
Cassinelli, but we modify it to an explosion instead of a steady
line-driven wind, we include a rotationally-distorted star, and we
treat the dynamics somewhat differently.  A continuum-driven
explosion, where the radiation force is independent of velocity,
avoids the disk inhibition mechanisms that normally operate in
line-driven winds.  This allows mid-latitude material with appropriate
initial specific energy to migrate toward the equator where it
collides with material from the opposite hemisphere to form a disk.
Thus, our model provides a simple method by which rotating hot stars
can simultaneously produce intrinsically bipolar and equatorial mass
ejections, without relying on an aspherical environment or magnetic
fields.  Although motivated by $\eta$ Carinae, the model may have
generic application to episodic mass ejection where rotation is
important, including other luminous blue variables, B[e] stars, the
nebula around SN1987A, or possibly even bipolar supernova explosions
themselves.  In cases where near-Eddington radiative driving is less
influential, our model generalizes to produce bipolar pinched-waist
morphologies without disks, as seen in many planetary nebulae.  If
rotating single stars can produce strongly axisymmetric ejecta by this
mechanism, then the presence of aspherical ejecta by itself is
insufficient justification to invoke close binary evolution.

\end{abstract}


\section{INTRODUCTION}

It is commonly assumed that shell nebulae surrounding massive hot
stars like luminous blue variables (LBVs) consist of slow ambient
material that has been swept-up by the faster wind of the hot
supergiant.  This scenario is often adopted to explain the origin of
bipolar geometry in their nebulae, by applying the generalized
interacting stellar winds (GISW) scenario that was developed
successfully for bipolar planetary nebulae and related outflow
phenomena (e.g., Mellema et al.\ 1991; Frank, Balick, \& Livio 1996;
Balick \& Frank 2002), as well as for the famous nebula around SN1987A
(Luo \& McCray 1991; Wang \& Mazzati 1992; Blondin \& Lundqvist 1993;
Martin \& Arnett 1995).  In this scenario, a hot fast wind expands
into a slower wind from a previous red supergiant (RSG) or asymptotic
giant branch (AGB) phase.  The surrounding slow wind must have an
equatorial density enhancement (i.e. a disk), and the consequent mass
loading near the equator slows the expansion of the shock interface
between the two winds, giving rise to a pinched waist and bipolar
structure.  However, it remains unclear how the required pre-existing
disk can be formed. One does not normally expect RSG or AGB stars to
be rotating rapidly enough to form an equatorial decretion disk such
as is characteristic to Be stars (Porter \& Rivinius 2003), and thus a
disk-shedding scenario probably requires the tidal influence of a
companion during prior evolutionary phases in order to add sufficient
angular momentum. In the case of SN~1987A, a binary merger would be
required for this particular scenario to work (Collins et al.\ 1999;
Morris \& Podsiadlowski 2006). As an alternative, variations of the
GISW paradigm can be invoked (e.g., including an aspherical fast wind
expanding into a spherical slow wind) to reproduce the bipolar shape
of the nebula around $\eta$~Carinae (Frank et al.\ 1995, 1998;
Dwarkadas \& Balick 1998; Langer et al.\ 1999; Gonzalez et al.\ 2004b,
2004b).


However, it is unlikely that this general scenario can work for
massive LBVs like $\eta$ Carinae, or for SN1987A.  Stars with
luminosities above roughly 10$^{5.8}$ L$_{\odot}$ never reach the RSG
stage, and the coolest apparent temperatures that they can achieve
occur instead during the LBV phase.\footnote{Admittedly, there is also
a group of relatively low luminosity LBVs around
log(L/L$_{\odot}$)=5.5 (see Smith, Vink, \& de Koter 2004a) where the
GISW may still apply, because these stars may be in a post-RSG phase.}
As they evolve off the main sequence, they move to the right on the HR
diagram, toward cooler temperatures, larger stellar radii, and lower
values for their escape velocities.  Consequently, their stellar wind
speeds get slower --- not faster --- as their mass-loss rates
increase.  O-type stars have typical wind speeds of a few 10$^3$ km
s$^{-1}$, whereas LBVs typically have terminal wind speeds of a few
10$^2$ km s$^{-1}$ and mass-loss rates a factor of $\sim$100-1000
higher.  This creates a situation where a slow dense wind is expanding
freely into a faster and much lower-density wind, which is {\it
exactly the opposite} situation of that required for the usual GISW
scenario to work.  In other words, the winds are not strongly
interacting.

The specific case of SN1987A presents its own set of difficulties,
even though its progenitor is well below log(L/L$_{\odot}$)=5.8 and it
probably has been through a recent RSG phase.  First, a merger model
followed by a transition from a RSG to BSG requires that these two
events be synchronized with the supernova event itself (to within the
$\sim$10$^4$ yr dynamical age of the nebula), requiring that the best
observed supernova in history also happens to be a very rare event.
One could easily argue, though, that the merger (needed for the
bipolar geometry) and the blue loop scenario might not have been
invented if SN1987A had occurred in a much more distant galaxy where
it would not have been so well-observed (i.e. we wouldn't know about
the bipolar nebula or the BSG progenitor).  Second, after the RSG
swallowed a companion star and then contracted to become a BSG, it
should have been rotating at or near its critical velocity (e.g.,
Eriguchi et al.\ 1992).  Even though pre-explosion spectra (Walborn et
al.\ 1989) do not have sufficient resolution to measure line profiles,
Sk--69$\arcdeg$202 showed no evidence of rapid rotation (e.g., like a
B[e] star spectrum).  Third, and particularly troublesome, is that
this merger and RSG/BSG transition {\it would need to occur
twice}. From an analysis of light echoes for up to 16 yr after the
supernova, Sugerman et al.\ (2005) have identified a much larger
bipolar nebula with the same axis orientation as the more famous inner
triple ring nebula.  If a merger and RSG/BSG transition are to blame
for the bipolarity in the triple-ring nebula, then what caused it in
the older one?  Perhaps a more natural explanation would be that
Sk--69$\arcdeg$202 suffered a few episodic mass ejections analogous to
LBV eruptions in its BSG phase (see Smith 2007).  The B[e] star R4 in
the Small Magellanic Cloud may offer a precedent at the same
luminosity as the progenitor of SN~1987A; R4 is consistent with a 20
M$_{\odot}$ evolutionary track, and it experienced a minor LBV
outburst in the late 1980's (Zickgraf et al.\ 1996).  R4 also has
elevated nitrogen abundances comparable to the nebula around SN~1987A.


For $\eta$ Carinae, observations have falsified the idea that the
bipolar shape arises from a prior equatorial density enhancement
(Smith 2006).  In the GISW scenario, the nebula's waist gets pinched
because of mass loading, so in the resulting bipolar nebula there
should be an excess of mass at low latitudes compared to a spherical
shell.  Instead, detailed observations of the latitudinal mass
dependence in the Homunculus show that the mass was concentrated
toward high polar latitudes (Smith 2006).  Furthermore, observations
show no evidence for a pre-existing slow disk around $\eta$ Car that
could have pinched the waist.  Instead, the disk structure that is
seen in high resolution images, usually called the ``equatorial
skirt'', has measured kinematics indicating an origin {\it at the same
time} as the bipolar lobes (Morse et al.\ 2001).  Some material
appears to be even {\it younger}, not older (Smith \& Gehrz 1998;
Davidson et al.\ 1997, 2001; Smith et al.\ 2004b; Dorland et al.\
2004).  Thus, any model for the production of the bipolar Homunculus
also needs to be able, simultaneously, to produce a thin equatorial
disk.  The only model proposed so far to accomplish this is in the
thermally-driven magnetic wind model of Matt \& Balick (2004) intended
for the presently-observed bipolar wind (Smith et al.\ 2003a).
However, it is not known if the extreme magnetic field required to
shape the massive Homunculus is achievable, because the conditions
during the 19th century eruption were much more extreme than in the
stellar wind seen now.


If LBV nebulae cannot be dominated by swept-up ambient material, we
are left in need of an alternative explanation for the origin of their
bipolar structure.  Several recent clues, due in large part to
detailed observations of $\eta$ Carinae, point instead toward the idea
that LBV eruptions can behave more like explosions than steady winds:

1.  The ratio of total mechanical energy to radiated energy for the
    19th century eruption of $\eta$ Car was greater than unity, and
    the ratio of kinetic momentum to photon momentum was $\sim$10$^3$
    (Smith et al.\ 2003b).  These numbers are characteristic of
    explosions rather than winds.

2.  Most of the mass in the Homunculus is concentrated in a very thin
    shell seen in H$_2$ emission (Smith 2006), which implies that
    $\Delta$t for the mass-loss event was $\la$5 yr.  Proper motions
    also suggest a similarly small range of ejection dates (Morse et
    al.\ 2001).  Ejecting the observed amount of mass in the
    Homunculus over that short a time requires a mass-loss rate of
    order 1 M$_{\odot}$ yr$^{-1}$ or higher (Smith et al.\ 2003b;
    Smith 2006), which may surpass the capability of even a
    super-Eddington continuum-driven wind by itself (Owocki et al.\
    2004).

3.  The high polar expansion speed in the Homunculus (about 650 km
    s$^{-1}$) is close to the expected escape velocity of the primary
    star ignoring radiation pressure.  In a radiatively-driven stellar
    wind, one expects the speed to drop as a star nears the Eddington
    limit because of the lower effective gravity. However, in an
    optically-thick hydrodynamic explosion that is not initially
    driven by radiation force, this condition would not necessarily
    apply.


In this paper we examine the idea that a surface explosion from a
rapidly-rotating hot star can create an intrinsically bipolar nebula
and thin disk, without relying on any latitude dependence in the
ambient material, binary influence, or magnetic fields.  Thus, our
model adopts the simplest set of assumptions that are also physically
plausible.

\begin{figure*}
\epsscale{0.7}
\plotone{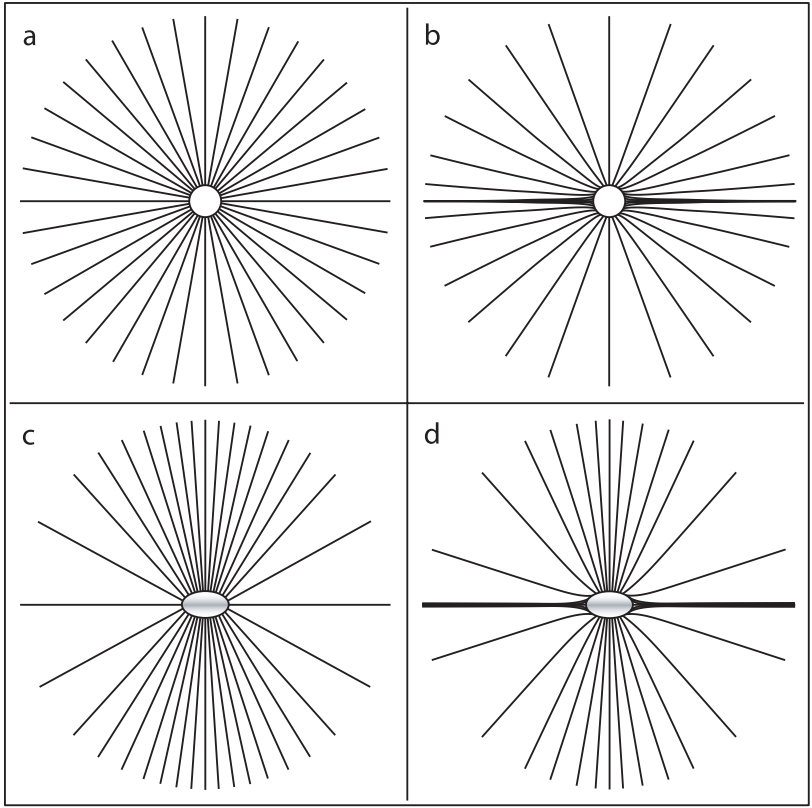}
\caption{A simple conceptual sketch of the various levels of
complexity for trajectories in a rotating explosion.  (a) Purely
radial trajectories from a spherical non-rotating star.  (b) Orbital
trajectories from a spherical rotating star, where material is
diverted toward the equator as in the wind compressed disk model of
Bjorkman \& Cassinelli (1993).  (c) Same as Panel $a$, but with
ejection normal to the surface of an oblate star.  (d) A combination
of Panels $b$ and $c$, where material is ejected normal to the surface
of an oblate star, but where rotation modifies the trajectories.  This
still enhances the polar density, but rapid rotation also diverts
material toward the equator to form a compressed disk.  The shading in
Panels $c$ and $d$ is to remind us of equatorial gravity darkening.}
\label{fig:sketch}
\end{figure*}

\section{THE MODEL: A ROTATING SURFACE EXPLOSION}

\subsection{Basic Principles}

Our simple model traces its origins back to a key question concerning
the morphology of $\eta$~Car's Homunculus: can a single basic physical
paradigm account for the simultaneous production of both the
equatorial skirt and the bipolar lobes? In considering this question
the phenomenon of rotation naturally springs to mind, since material
kicked ballistically from the surface of a rotating star has a natural
tendency to migrate toward the stellar equatorial plane. This notion
was invoked in the wind-compressed disk (WCD) model of Bjorkman \&
Cassinelli (1993) as a means for explaining the circumstellar disks
around Be stars. The WCD model treats the wind plasma as independent
fluid parcels whose dynamics are confined to their respective orbital
planes.  Detailed calculations by Owocki, Cranmer \& Gayley (1998)
demonstrated that subtle line-driving effects associated with
velocity-dependent forces tend to inhibit the formation of disks in
the WCD model. However, in our treatment below we focus on continuum
driving, which is independent of velocity, and the WCD model furnishes
a useful paradigm on which to build.

We advance the hypothesis that, to first order, the shape of the
Homunculus can be understood as the consequence of the nearly
ballistic, anisotropic ejection of the surface layers of a star near
its critical rotation limit.  In this scenario, the bipolar shape and
equatorial disk arise as a direct consequence of the ejection
geometry, rather than through hydrodynamic interaction afterward.  We
assume that the subsequent evolution of these layers is governed
solely by a central, gravity-like (1/$r^2$) force, and therefore that
the trajectory followed by each fluid element may be treated as a
problem in orbital dynamics.  The exception is for elements that pass
through the equatorial plane; there, we assume that they collide with
material from the opposite hemisphere, leading to the cancellation of
their velocity component normal to the plane.  We ignore the self
gravity of the ejected mass.

Sketches of the trajectories for outflowing material under various
levels of complexity are illustrated in Figure 1.  These trajectories
show in principle how density can be enhanced at the equator to form a
disk, as well as toward the poles.  When a spherical star with radial
trajectories (Fig.\ 1$a$) is allowed to rotate rapidly, material is
diverted toward the equator (Fig.\ 1$b$) by conservation of angular
momentum, as envisioned in the WCD model.  However, a severely oblate
star introduces other effects as well.  First, as noted above, an
oblate star has a smaller radius, larger effective gravity, and larger
escape velocity at the pole than at the equator.  This -- rather than
the latitude dependence of density -- is what gives rise to the
bipolar {\it shape} of the nebula.  The oblateness of the star will
also enhance the polar {\it density} if material is initially launched
normal to the surface, simply because the flatter poles of the star
aim trajectories poleward (Fig.\ 1$c$).  We do not attempt to model
this enhanced polar density quantitatively, however, because we expect
that additional effects may be as important.  For example, the polar
mass flux can be increased by the stronger radiative flux at the poles
of a star that suffers from rotationally-induced gravity darkening
(e.g., Owocki \& Gayley 1997; Owocki et al.\ 1998).  In any case, the
straight (i.e. non-rotating) trajectories from a
rotationally-distorted star in Figure 1$c$ seem unrealistic unless the
ejecta speeds are much faster than the rotation speed.  The more
likely rotationally-modified trajectories are shown conceptually in
Figure 1$d$.  This last case creates a compressed disk, but also
retains some degree of enhanced density at the poles, because the
rotational wind compression effects are more important at larger
cylindrical radii.

Elaborating on this sketch, we consider a model star in an
initial state of critical rotation. (We have no direct evidence that
$\eta$~Car was rotating critically prior to the Great Eruption;
however, since the investigation of rotational effects is central to
our study, it is appropriate to focus on the critical limit in which
these effects are the most pronounced.)  In the Roche approximation
(e.g., Cranmer 1996, and references therein), the surface radius of
the oblate star is given by
\begin{equation} \label{eqn:roche}
R(\theta) = 
\begin{cases}
3 \frac{\sin (\theta/3)}{\sin\theta}\, R_{\rm p} & \theta < \pi/2 \\
R(\pi - \theta)  & \theta > \pi/2 \\
\end{cases}
\end{equation}
where $\theta$ is the usual polar coordinate, and $R_{\rm p}$ is the
polar radius of the star. The azimuthal velocity due to rotation is
given by
\begin{equation} \label{eqn:v-rot}
V_{\phi}(\theta) = \Omega_{\rm c} R(\theta) \sin\theta,
\end{equation}
where
\begin{equation}
\Omega_{\rm c} \equiv \sqrt{\frac{8 G M}{27 R_{\rm p}^{3}}},
\end{equation}
is the critical angular frequency of rotation in the same Roche
approximation, with $M$ the stellar mass and $G$ the gravitational
constant.

At $t=0$, we disrupt this initial state by (i) imparting a
velocity kick $V_{\rm k}(\theta)$ to each surface fluid element in the
direction of the local surface normal (this is the {\it explosion}),
and (ii) introducing a spherically symmetric force that everywhere is
directed radially outward, and whose magnitude is $\alpha$ times the
gravitational force. (It is assumed that $\alpha < 1$, so that
the \emph{net} force on elements remains directed inward toward the
star). This disruption represents our basic characterization of the
Great Eruption observed in the 1840's, when the nebular material is
thought to have been launched (Morse et al.\ 2001; Smith \& Gehrz
1998; Currie et al.\ 1996; Gehrz \& Ney 1972; Ringuelet 1958; Gaviola
1950).  For simplicity we assume instantaneous ejection rather than
sustained outflow over 10-20 years, but this should have little effect
on the overall results.  A series of bursts over a decade, as opposed
to one single burst, would produce a nebula with a similar shape but
some finite thickness.

Our motivation for including the kick (i) and the subsequent
outward force (ii) comes from considering the effects of continuum
radiation driving during the sudden, factor $\sim 5$ increase in
luminosity associated with the Great Eruption (see, e.g., Davidson \&
Humphreys 1997). At the beginning of the eruption, rotation-induced
gravity darkening will produce a strongly anisotropic radiation field
(e.g., von Zeipel 1924; Cranmer \& Owocki 1995; Owocki et al.\ 1996,
1998; Owocki \& Gayley 1997; Langer 1998; Glatzel 1998; Maeder 1999;
Maeder \& Desjacques 2001). This means that the additional radiative
flux escaping the star will at first deposit momentum preferentially
at the stellar poles; by assuming an appropriate form for $V_{\rm
k}(\theta)$ (discussed in greater detail in the following sections),
we use the velocity kick (i) to model this initial polar
deposition. As the surface fluid elements subsequently move outward,
however, the anisotropies in the radiative flux will tend to be
smeared out, leading to a more spherically-symmetric outward radiative
force that we incorporate via (ii).

To determine the $t>0$ evolution of the surface fluid
elements, we assume that each follows a trajectory described by the
equation of motion
\begin{equation} \label{eqn:motion}
\frac{{\rm d} \mathbf{v}}{{\rm d} t} = - \frac{G M}{r^{3}}\mathbf{r} +
  \alpha \frac{G M}{r^{3}} \,\mathbf{r}.
\end{equation}
Here, $\mathbf{r}$ is the position vector of the element, $\mathbf{v}
\equiv {\rm d}\mathbf{r}/{\rm d}t$ the corresponding velocity vector,
and we adopt the convention that non-bold symbols denote the modulus
of their bold vector equivalents, so that in this case $r \equiv
|\mathbf{r}|$. The acceleration terms on the right-hand side of this
equation arise, respectively, from the inward gravitational force and
the outward spherically-symmetric force introduced at $t=0$. Following
the discussion given above, the initial velocity of each element is
calculated as
\begin{equation} \label{eqn:v-initial}
\mathbf{v}_{0} = V_{\rm k}(\theta) \,\mathbf{e}_{n} + V_{\phi}(\theta)
\,\mathbf{e}_{\phi}
\end{equation}
where
\begin{equation}
\mathrm{e}_{n} = \frac{\mathbf{e}_{r} - (R'/R)
  \,\mathbf{e}_{\theta}}{\sqrt{1 + (R'/R)^{2}}}
\end{equation}
is the unit surface normal vector, with $R' \equiv \partial
R/\partial\theta$, and
$\{\mathbf{e}_{r},\mathbf{e}_{\theta},\mathbf{e}_{\phi}\}$ are the
unit basis vectors in the spherical-polar $\{r,\theta,\phi\}$
directions.

The equation of motion~(\ref{eqn:motion}) is identical to that
for a test particle moving in the gravitational field of a point mass
$M(1-\alpha) > 0$. Therefore, the solutions are analytical, taking the
form of conic sections (ellipses, hyperbolae, etc.) whose focus lies
at the stellar origin. We leave a detailed discussion of these
solutions to any of the many standard texts discussing this classical
two-body problem (e.g., Boccaletti \& Pucacco 1996). However, it is
appropriate to specify how the six orbital parameters, defining the
trajectory followed by each surface fluid element, are determined from
the initial conditions. First, we calculate the specific
(per-unit-mass) angular momentum vector
\begin{equation}
\mathbf{j} = \mathbf{r}_{0} \times \mathbf{v}_{0},
\end{equation}
(where $\mathbf{r}_{0}$ is the element's position at $t=0$, and
$\mathbf{v}_{0}$ was defined in eqn.~(\ref{eqn:v-initial})), and the
specific Laplace-Runge-Lenz vector
\begin{equation}
\mathbf{A} = \mathbf{v} \times \mathbf{j} - GM (1 - \alpha) \mathbf{e}_{r}.
\end{equation}
We assume that the reference plane is the stellar equatorial
(Cartesian $x-y$) plane, with the $x$-axis defining the Vernal
point. Then, the inclination $i$ of the orbital plane is given by
\begin{equation}
\cos i = \frac{\mathbf{j} \cdot \mathbf{e}_{z}}{j};
\end{equation}
the longitude of the ascending node $\Omega$ and argument of periastron
$\omega$ by
\begin{equation}
\tan \Omega = \frac{\mathbf{e}_{z} \cdot [\mathbf{e}_{x} \times
    (\mathbf{e}_{z} \times \mathbf{j})]}{\mathbf{e}_{x} \cdot
    (\mathbf{e}_{z} \times \mathbf{j})},
\qquad
\tan \omega = \frac{\mathbf{j} \cdot [\mathbf{A} \times
    (\mathbf{e}_{z} \times \mathbf{j})]}{j \mathbf{A} \cdot
    (\mathbf{e}_{z} \times \mathbf{j})};
\end{equation}
and the eccentricity $e$ and semi-major axis $a$ by
\begin{equation}
e = \frac{A}{GM(1 - \alpha)},
\qquad
a = \left| \frac{GM(1 - \alpha)}{v^{2} - 2GM(1-\alpha)/r} \right|.
\end{equation}
(In these expressions, $\{\mathbf{e}_{x},\mathbf{e}_{z}\}$ are the unit
basis vectors in the Cartesian$\{x,z\}$ directions.) Finally, the true
anomaly $\upsilon_{0}$ of the surface element at the $t=0$ epoch is
given by
\begin{equation}
\tan \upsilon_{0} = \frac{\mathbf{j} \cdot (\mathbf{A} \times
  \mathbf{r}_{0})}{j \mathbf{A} \cdot \mathbf{r}_{0}}.
\end{equation}

The orbital parameters $\{i,\Omega,\omega,e,a,\upsilon_{0}\}$ defined
above allow calculation of the complete $t>0$ evolution of a given
surface element. However, for those elements whose trajectories pass
through the equatorial plane, the parameters must be modified to
account for the anticipated collision with material from the opposite
hemisphere. As discussed above, the polar ($\theta$) component of the
velocity $\mathbf{v}$ is set to zero when the element reaches the
equator. Then, the orbital parameters are recalculated using the
element's instantaneous position and updated velocity, and its
evolution is continued.

This approach is rather different from the WCD model of Bjorkman \&
Cassinelli (1993), who assumed that velocity vectors become radial at
the equator, with no change in speed. In fact, a more fundamental
difference between our approach and the WCD model lies in the
treatment of the outward force introduced at $t=0$. Bjorkman \&
Cassinelli (1993) incorporated a parameterization of line-driven wind
theory in their model, resulting in an outward radiative force that
(a) exceeds gravity, and (b) does not have a simple $1/r^{2}$
scaling. The significance of (a) is that an initial kick (as assumed
in our treatment) is not required for material at the stellar poles to
escape from the star. However, (b) means that the equation of motion
does not correspond to a two-body gravitational problem, and must be
integrated numerically.

\begin{figure*}
\epsscale{0.99}
\plotone{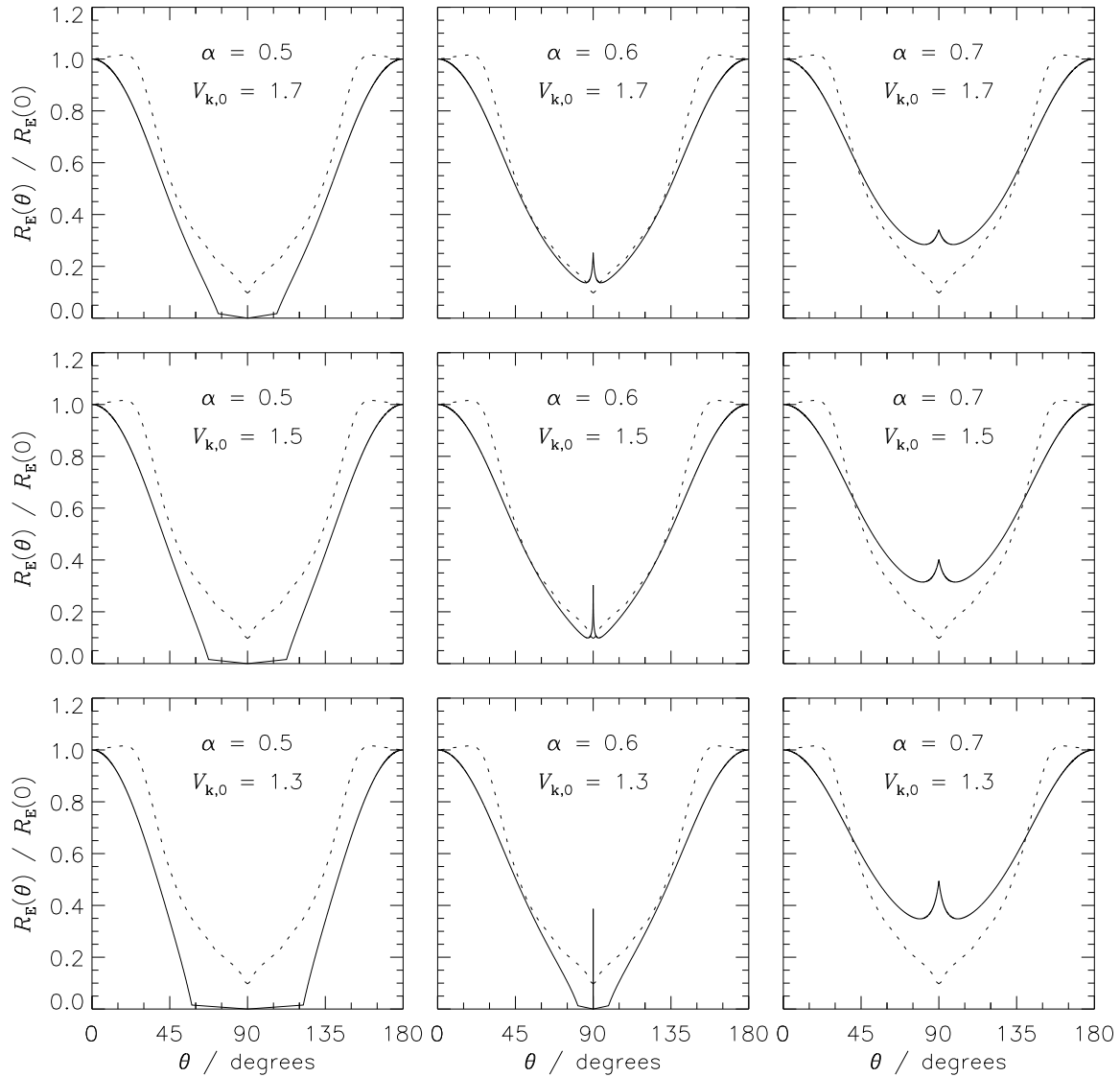}
\caption{The envelopes $R_{\rm E}(\theta)$ of the ejected surface
elements in the asymptotic limit $t \rightarrow \infty$ (solid),
plotted together with the measured shape of the Homunculus (dotted)
from Smith (2006).  Each panel shows results from a simulation with
the indicated gravity reduction parameter $\alpha$ and velocity kick
normalization $V_{k,0}$, the latter being measured in units of
$\sqrt{GM/R_{\rm p}}$.}
\label{fig:demo}
\end{figure*}

\subsection{Illustrative Simulations}

To furnish an initial demonstration of our approach, we consider the
case where the velocity kick function is given by
\begin{equation} \label{eqn:v-kick}
V_{\rm k}(\theta) = V_{{\rm k}, 0} |\cos\theta|,
\end{equation}
for some normalizing velocity $V_{{\rm k},0}$ at the poles, which in
the case of $\eta$ Car is about 650 km s$^{-1}$ (Smith 2006).  The
dependence on $|\cos\theta|$ may seem {\it ad hoc} in our simulation,
because it essentially prescribes the overall shape that is observed.
However, this latitude dependence for the initial kick has a firm
physical justification. \textbf{As we have discussed above, gravity
darkening --- whereby the local emergent flux scales with the local
effective gravity --- tends to initially focus the additional
radiative flux escaping from the star toward the stellar poles. Thus,
the kick imparted by this flux is expected to be strongest over the
poles, suggesting the above form for $V_{\rm k}(\theta)$.}

We should mention that a prescription similar to~(\ref{eqn:v-kick})
above has already been invoked to explain the bipolar shape of the
Homunculus Nebula and the latitude dependence of $\eta$~Car's stellar
wind for near-critical rotation (Owocki 2005; see also Owocki \&
Gayley 1997; Maeder \& Desjacques 2001; Dwarkadas \& Owocki 2002;
Smith 2002, 2006; Smith et al.\ 2003a). However, the focus in most of
these studies is on the terminal velocity $v_{\infty}$ of a
line-driven wind, whereas in the present study we are considering the
initial kick velocity of an explosive ejection. The reason why a
$\cos\theta$ variation is appropriate in the line driven case is that
$v_{\infty}$ typically scales with the local escape velocity $v_{\rm
esc}$ (see, e.g., Dwarkadas \& Owocki 2002). Then, with $v_{\rm esc}$
itself scaling with effective gravity in the same way as the radiative
flux, a coincidence between $V_{\rm k}$ and the line-driven
$v_{\infty}$ naturally arises.

For a selected region of $\alpha$--$V_{{\rm k},0}$ parameter space, we
conducted simulations where we evolve a set of surface elements to the
asymptotic limit $t \rightarrow \infty$. For each simulation, the
initial state at $t=0$ is comprised of 5,000 elements distributed
uniformly in $\theta$ over the Roche surface described by
eqn.~(\ref{eqn:roche}). With initial velocities described by
eqns.~(\ref{eqn:v-rot},\ref{eqn:v-initial},\ref{eqn:v-kick}), these
elements are evolved according to the equation of
motion~(\ref{eqn:motion}), as described in the preceding section. In
the limit of large $t$, the envelope $R_{\rm E}(\theta)$ defined by
the elements reaches a steady state that corresponds to the eventual
shape of the ejected nebula.

\begin{figure}
\epsscale{0.95}
\plotone{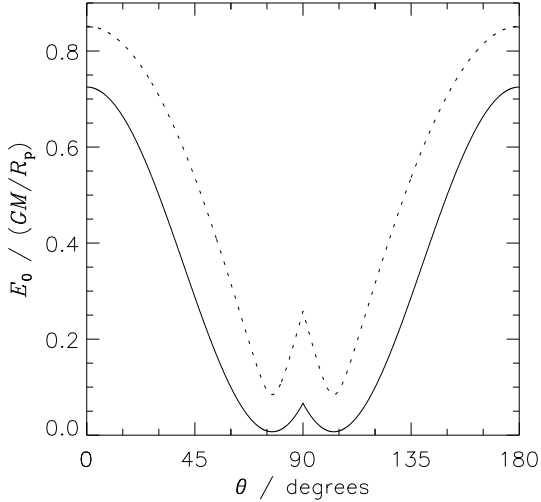}
\caption{The initial specific (per-unit-mass) energy $E_{0}$ (solid),
and its square root $\sqrt{E_{0}}$ (dotted), plotted as a function of
$\theta$ for the $\alpha=0.6$, $V_{k,0} = 1.5$ simulation displayed in
Fig.~\ref{fig:demo}. }
\label{fig:energy}
\end{figure}

Figure~\ref{fig:demo} compares simulated envelopes $R_{\rm E}(\theta)$
against the shape of the $\eta$ Car Homunculus as measured by Smith
(2006). Clearly, with an appropriate choice of parameters --- in this
case, $\alpha = 0.6$ and $V_{{\rm k},0} = 1.5\,\sqrt{GM/R_{\rm p}}$
--- we are able to capture the gross qualitative features of the
bipolar nebula, while at the same time producing the desired
equatorial skirt. Generally, a skirt occurs whenever the initial
specific energy of surface elements,
\begin{equation}
E_{0} = |\mathbf{v}_{0}|^{2}/2 - GM(1 - \alpha)/r_{0},
\end{equation}
exhibits a minimum at some point between equator and pole. (Such
minima themselves arise because the kick kinetic energy decreases
toward the equator, while the rotational kinetic energy and
gravitational potential energy both increase toward the equator.)  The
surface elements situated at the energy minimum lead to the narrow
waist of the envelope. The elements closer to the equator, with higher
$E_{0}$ and hence faster terminal velocities $v_{\infty} \propto
\sqrt{E_{0}}$, then form the skirt, while the elements closer to the
poles produce the bipolar lobes. In Fig.~\ref{fig:energy} we plot both
$E_{0}$ and $\sqrt{E_{0}}$ as a function of $\theta$ for the $\alpha =
0.6$, $V_{{\rm k},0} = 1.5\,\sqrt{GM/R_{\rm p}}$ simulation. The
specific energy minima that give rise to the skirt in this case can
clearly be seen at $\theta \approx 80^{\circ}$ and $\theta \approx
100^{\circ}$. Note that the tendency of elements to evolve toward the
equatorial plane, due to conservation of angular momentum, produces a
rather narrower skirt than might be assumed from a cursory look at the
plot; compare, for instance, the $\sqrt{E_{0}}$ data shown in
Fig.~\ref{fig:energy}, with the center panel of Fig.~\ref{fig:demo}.
Figure 3 reproduces the trend of higher kinetic energy in the polar
ejecta observed in the Homunculus (Smith 2006).

Toward smaller values of $\alpha$ and/or $V_{{\rm k}, 0}$ the initial
specific energy of surface elements near the equator is
negative. These elements therefore remain gravitationally bound to the
star, producing a collapsed-waist morphology with no skirt.  The
implication is that, when generalized to cases where near-Eddington
radiative driving is less influential ($\alpha\la$0.5 in Fig.\ 2), our
model simplifies and easily accomodates the more common pinched waist
morphologies of planetary nebulae without fast skirts (e.g., Balick \&
Frank 2002).  This is encouraging, since these sources have central
stars that are indeed far less luminous than $\eta$ Carinae.
Conversely, toward larger $\alpha$ the elements have insufficient
angular momentum to reach the equatorial plane, and the skirt is
replaced by a local inversion of the bipolar shape.  This range of
conditions may explain why equatorial disks like that around
$\eta$~Car are not always seen, while bipolar ejecta nebulae are quite
common around evolved rotating stars.

\subsection{A Tuned Simulation}

A notable discrepancy between observations and our `best' (central
panel) simulation shown in Figure~\ref{fig:demo} is that the apparent
flattening of the nebula lobes over the poles is not properly
reproduced.  This can be seen as faster expansion in the Homunculus at
latitudes about 15--40\arcdeg\ from the polar axis, as compared to our
predicted shape.  It is as if the Homunculus has received an extra
kick at these latitudes.

\begin{figure*}
\epsscale{1.05}
\plotone{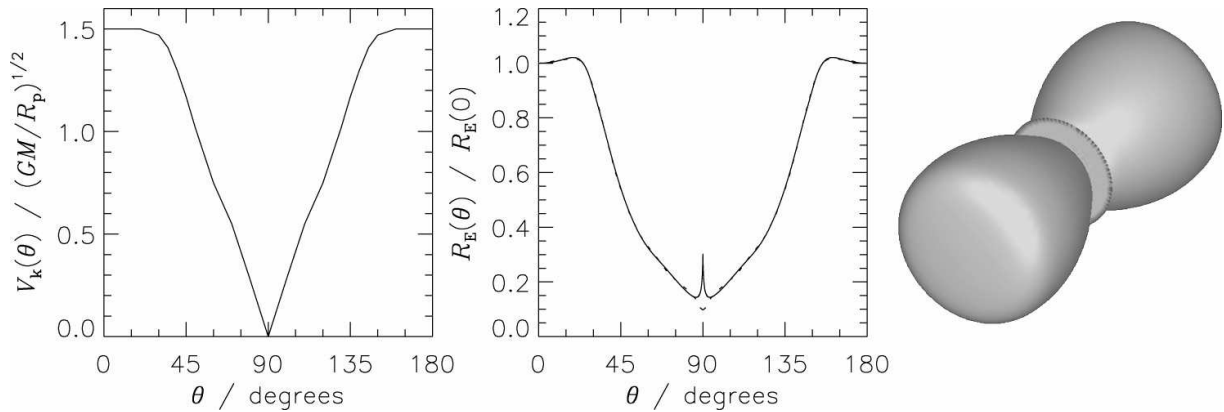}
\caption{The initial kick velocity function (left panel), in units of
$\sqrt{GM/R_{\rm p}}$, tuned to reproduce the shape of the Homunculus
(center panel; solid and dotted lines have the same meaning as in
Fig.~\ref{fig:demo}). A 3 dimensional rendering of the corresponding
surface of revolution is shown in the right panel; the inclination
($i=41^{\circ}$) and orientation (${\rm P.A.}  = 310^{\circ}$) of the
surface are based on the values published by Smith
(2006).}\label{fig:tuned}
\end{figure*}

What $V_{\rm k}(\theta)$ must be adopted in order to reproduce
correctly the lobe shapes?  One possible answer to this question is
presented in Figure~\ref{fig:tuned}.  Here, the kick velocity function
has been tuned with the specific purpose of reproducing the lobe
shapes. The tuning was accomplished through a simple process of trial
and error, subject to the constraints that $\alpha = 0.6$ and that
over the stellar poles $V_{\rm k} = 1.5\,\sqrt{GM/R_{\rm p}}$ (these
values come from the `best' simulation of the preceding section). The
close match between theory and observations seen in the central panel
of Figure~\ref{fig:tuned} should not be taken as a measure of the
fidelity of our model --- indeed, since we have adopted an \emph{ad
hoc} prescription for $V_{\rm k}(\theta)$, the close agreement is to
be expected.

However, what \emph{is} of particular significance is the fact that
the simulation simultaneously reproduced the lobes \emph{and} a skirt
that is similar to that which is actually seen around $\eta$~Car
(e.g., Duschl et al.\ 1995).  Namely, this is a true flattened
disk-like structure, with material of the same age spread out over a
range of radii in the equatorial plane -- it is not a ring, even
though we adopted an instantaneous explosion event (it is also
kinematically different from a Keplerian disk).  This is a true
success of our model, since the signature of this skirt is wholly
absent from the initial kick velocity function in the left panel of
Figure~3.

Furthermore, the small extra kick needed at latitudes about
15--40\arcdeg\ from the polar axis may have a reasonable physical
explanation.  Although our model attributes the overall shape of the
Homunculus and its disk to initial conditions of the ejection, this is
a case where hydrodynamic shaping of the ejecta long after ejection
may play some role after all.  We have argued that the GISW scenario
cannot drive the overall shape of the Homunculus, largely because the
post-eruption wind is not powerful enough (Smith 2006; Smith et al.\
2003a,b).  However, the post-eruption wind may be able to modify or
perturb the existing shape.  It is often seen in hydrodynamic
simulations of interacting winds that bipolar nebulae develop
``corners'' at the outer parts of the polar lobes, leading to
flattening over the poles that is reminiscent of the extra velocity we
require here (e.g., Cunningham et al.\ 2005; Frank et al.\ 1998;
Dwarkadas \& Balick 1998).  This effect arises in the following way:
at low latitudes within, say, 45\arcdeg\ of the equator, the fast wind
that is inflating the bipolar nebula strikes the inner side walls of
the dense polar lobes at an oblique angle, and is then deflected
poleward.  This material skims along the inner side walls of the polar
lobes and piles up in the corners (see Cunningham et al.\ 2005),
adding an extra kick over a small range of latitudes.  Although the
average momentum of $\eta$~Car's post eruption wind is tiny compared
to the momentum of the nebula (Smith et al.\ 2003b), in this scenario,
the momentum from a relatively large volume fraction of the
post-eruption wind is focused on only a small portion of the nebula.
This effect is purely hydrodynamic, so our {\it ad hoc} approach in
Figure 3 may not be the best way to explore it.  A better (but
computationally more intense) way might be to take our predicted model
shape in the central panel of Figure 2, and allow it to be
``inflated'' and shaped self-consistently by a post-eruption wind.

The potential role of the post-eruption stellar wind (which is a
line-driven wind) that follows the continuum-driven outburst also
suggests an interesting ``double whammy'' for enhancing the bipolar
shapes of LBV nebulae.  Specifically, Owocki and collaborators (Owocki
\& Gayley 1997; Owocki et al.\ 1996, 1998) have demonstrated that
radiative line driving not only inhibits the initial formation of an
equatorial disk, but also enhances the wind mass-loss rate toward the
stellar poles.  In the case of $\eta$ Car, this polar wind has been
observed (Smith et al.\ 2003a).  Such a polar wind may enhance the
bipolar shape of the initial ejection.

Of course, one can imagine other factors that may modify the shape of
the polar lobes in the required way, such as the influence of a nearby
companion star (e.g., Morris \& Podsiadlowski 2006).  This realm of
shaping mechanisms is beyond the scope of our present study, but
should be explored further.

\section{DISCUSSION}

Using a semi-analytic model, we have shown that a surface explosion
from an oblate star near critical rotation can simultaneously produce
an equatorial disk and a pair of polar lobes that closely approximate
the observed shape of the Homunculus Nebula around $\eta$ Carinae.
This model is arguably the simplest model that also includes realistic
assumptions.  It does not require any effects of hydrodynamically
interacting winds or magnetic fields to produce the asymmetry.  This
shows that rotating hot stars can eject intrinsically bipolar nebulae
simultaneously with equatorial disks.

Our techniques combine two aspects of theories developed initially for
non-spherical line-driven stellar winds, but we modify them to the
scenario of a sudden explosion.  Namely, our model borrows
conceptually from the WCD model of Bjorkman \& Cassinelli (1993), as
well as the expectation that the ejection speed is proportional to the
latitudinal variation of the escape speed on the surface of a rotating
star, as noted by several investigators.  By adapting these ideas to
an explosion with continuum driving, however, our model does not
suffer from the difficult problem of WCD inhibition due to effects
associated with non-radial forces in line-driven winds (Owocki et al.\
1998).  Whether or not our model is applicable depends on the nature
of the episodic mass-loss event in any individual case; the mass loss
must be strong enough that it is not dominated by a line-driven wind
during an outburst (see Smith \& Owocki 2006; Owocki et al.\ 2004).

Previous models to explain the shapes of bipolar nebulae around $\eta$
Carinae and other LBVs differ substantially from ours.  Most
approaches have used hydrodynamic simulations of interacting winds,
where a fast wind sweeps into an equatorial density enhancement to
produce the bipolar shape, or variations of that scenario (Frank et
al.\ 1995, 1998; Dwarkadas \& Balick 1998; Langer et al.\ 1999;
Gonzalez et al.\ 2004a, 2004b).  However, none of these produced an
equatorial disk with the same apparent age as the polar lobes.  Soker
(2004; and references therein) has discussed a complex model where
accretion onto a companion star drives jets that shape the bipolar
lobes, much as in similar models for planetary nebulae, but this model
also fails to account for the equatorial disk.  Other approaches
involving non-spherical stellar winds are closer to our own, where the
bipolar shape is an intrinsic feature of ejection from a rotating star
(Owocki 2003, 2005; Owocki \& Gayley 1997; Dwarkadas \& Owocki 2002;
Maeder \& Desjacques 2001).  In these models, the shape of the
Homunculus is achieved by such a wind blowing with an enhanced
mass-loss rate for a short time.  Once again, however, none of these
produced an equatorial disk.\footnote{Maeder \& Desjacques (2001)
presented a second case where a steady stellar wind included a dense
disk.  However, while this model may account for the enhanced density
at the equator, it does not account for the shape (i.e. the speed)
that can produce a disk at the same time as the polar lobes.  It also
depends on effects in line-driven winds that are not applicable to the
Great Eruption of $\eta$ Carinae that produced the Homunculus; S.\
Owocki (private comm.) has noted the difficulty in forming disks via
this type of opacity mechanism in radiatively-driven winds.}

One model that did simultaneously produce bipolar lobes and a disk was
the thermally-driven magnetohydrodynamic rotating stellar wind model
of Matt \& Balick (2004).  While that model was encouraging, we felt
it was also useful to pursue alternatives that did not rely on strong
magnetic fields, since it is unclear whether the huge fields required
to shape the 10--15 M$_{\odot}$ (Smith et al.\ 2003b) ejected during
the Great Eruption are achievable.

Two main assumptions required to produce the observed shape are
explosive mass loss with continuum radiative driving and near-critical
rotation.  Explosive mass loss is justified by the observed fact that
$\eta$ Car lost a huge amount of mass in a short time, with a required
mass-loss rate that is too high to be accounted for with a line-driven
driven wind, as noted earlier.  Some recent stellar evolution models
for very massive stars (Arnett et al.\ 2005; Young 2005) predict
deep-seated hydrodynamic explosions that can potentially release the
required amount of kinetic energy of almost 10$^{50}$ ergs seen in the
Homunculus (Smith et al.\ 2003b).  Near critical rotation is expected
to occur at late evolutionary stages as a BSG (e.g., Langer et al.\
1999; Eriguchi et al.\ 1992).  Soker (2004) has criticized single-star
models for the ejection and shaping of the Homunculus based on the
idea that the star spun down during the eruption and was not rotating
sufficiently rapidly.  However, two points of clarification should be
mentioned in this regard.  First, near the classical Eddington limit,
a star can be near critical rotation for mass loss without necessarily
being a true rapid rotator because of the effects of radiation
pressure.  Second, while the star may have shed angular momentum by
ejecting the Homunculus and therefore was rotating more slowly after
the outburst, it is the {\it initial} rotation and available angular
momentum at the time of ejection that is critical, not the end state,
especially in an explosion scenario such as that we are discussing
here.

The simultaneous formation of a disk and bipolar lobes from the same
mechanism is a novel feature of our model, and there are interesting
implications for the observed structures.  The equatorial ejecta of
$\eta$~Carinae were originally described as a ``skirt'' rather than a
``disk'' because of their ragged, streaked appearance, which is
different from the clumpy character of the polar lobes (e.g., Morse et
al.\ 1998; Duschl et al.\ 1995; Zethson et al.\ 1999). This expanding
excretion structure should not be confused with a Keplerian disk.  In
our proposed scenario for the formation of $\eta$ Car's skirt,
material ejected from mid latitudes on the star migrates toward the
equatorial plane, where it crashes into material from the opposite
hemisphere.  One can easily envision a violent collision that could
lead to the ragged spray of ejecta seen today, rather than a smooth
disk, and hydrodynamic simulations of this would be interesting.  It
seems likely that this violent splashing at the equatorial plane could
create even faster disk material than the speeds seen in our
simulations where we simply canceled-out the vertical component of the
velocity.  To be sure, there are additional complexities in
$\eta$~Car's equatorial ejecta that we do not even attempt to treat
here.  Most notable among them are the origin of the NN jet (Meaburn
et al.\ 1993; Walborn \& Blanco 1988) and the presence of younger
ejecta from the 1890 eruption that appear to co-exist with older
ejecta in the skirt (Smith \& Gehrz 1998; Davidson et al.\ 2001; Smith
et al.\ 2004b).

Finally, we expect that our proposed scenario could have wider
applications beyond the Homunculus around $\eta$~Carinae.  A simple
mechanism for how a rotating hot star can simultaneously produce
equatorial and polar ejecta might be relevant for the famous
triple-ring nebula around SN1987A (Burrows et al.\ 1995), providing a
possible way to circumvent difficulties in explaining this object via
the GISW scenario (see Smith 2007).  Likewise, it may apply to bipolar
ejecta and rings around other blue supergiants in our galaxy such as
Sher 25 (Brandner et al.\ 1997) and the LBV candidate HD168625 (Smith
2007), as well as LBVs in general.  If extragalactic analogs of $\eta$
Carinae behave similarly in their outbursts, then one might expect the
so-called ``supernova impostors'' to be significantly polarized.  The
mechanism may also be applicable to the short lived emitting disks
around B[e] stars (Zickgraf et al.\ 1986, 1996), or possibly other hot
stars where episodic mass ejection is important.

An encouraging property of our model is that in cases where
near-Eddington radiative driving is less influential than in $\eta$
Carinae (such as in lower-luminosity planetary nebulae), the mechanism
proposed here generalizes to a situation that reproduces a simple
bipolar pinched-waist morphology without an obvious fast disk.  Thus,
scenarios like those in the left column of Figure 2 may have wide
application to observed morphologies of bipolar planetary nebulae,
which generally lack such disks (Balick \& Frank 2002).

Some supernova explosions are seen to be intrinsically asymmetric; the
bipolar supernova ejecta in SN1987A are seen directly (Wang et al.\
2002), while others show polarization at early times (e.g., Leonard et
al.\ 2000, 2001; Leonard \& Filippenko 2001).  If near-critical
rotation is important at some point within these explosions, it is
conceivable that our model described here may offer an alternative to
jet-driven hydrodynamic models for explaining some aspects of
asymmetric core-collapse supernovae.

In any case, a viable mechanism for a single star to produce strongly
axisymmetric ejecta means that the presence of asymmetry in the
circumstellar environment is, by itself, not a valid justification to
invoke close binary interactions in a supernova progenitor or any
other hot massive star.

\acknowledgments \scriptsize

We gratefully acknowledge many fruitful discussions and collaborations
with Stan Owocki, which have shaped our view and aided our
understanding of non-spherical mass loss, and we thank an anonymous
referee for constructive comments.  N.S.\ was partially supported by
NASA through grant HF-01166.01A from the Space Telescope Science
Institute, which is operated by the Association of Universities for
Research in Astronomy, Inc., under NASA contract NAS5-26555. R.H.D.T.\
was supported by NASA grant NNG05GC36G.


\end{document}